\documentclass{elsart}

\usepackage{graphicx}% Include figure files

\begin{document}

\begin{frontmatter}

\title{Bell's inequality without alternative settings}
\author{Ad\'{a}n Cabello}
%\email{adan@us.es}
\ead{adan@us.es}
%\affiliation{Departamento de F\'{\i}sica Aplicada II, Universidad de Sevilla, 41012 Sevilla, Spain}
\address{Departamento de F\'{\i}sica Aplicada II, Universidad de Sevilla, 41012 Sevilla, Spain}
%\date{\today}
%First version: July 19, 2002.
%This version: May 9, 2003.
%After PLA's proofs.

%%%%%%%%%%%%%%%%%%%%%%%%%%%%%%%%%% Abstract %%%%%%%%%%%%%%%%%%%%%%%%%%%%%%%%%%%

\begin{abstract}
A suitable generalized measurement described by a 4-element
positive operator-valued measure (POVM) on each particle of a
two-qubit system in the singlet state is, from the point of view
of Einstein, Podolsky, and Rosen's (EPR's) criterion of elements
of reality, equivalent to a random selection between two
alternative projective measurements. It is shown that an
EPR-experiment with a fixed POVM on each particle provides a
violation of Bell's inequality without requiring local observers
to choose between the alternatives. This approach could be useful
for designing a loophole-free test of Bell's inequality.
\end{abstract}

%%%%%%%%%%%%%%%%%%%%%%%%%%%%%%%%%%%%%%%%%%%%%%%%%%%%%%%%%%%%%%%%%%%%%%%%%%%%%%%

\end{frontmatter}

%\pacs{03.65.Ud,
%Entanglement and quantum nonlocality
%(e.g. EPR paradox, Bell's inequalities, GHZ states, etc.)
%03.65.Ta}
%Foundations of quantum mechanics; measurement theory
%\maketitle

%%%%%%%%%%%%%%%%%%%%%%%%%%%%%%%%%%%%%%%%%%%%%%%%%%%%%%%%%%%%%%%%%%%%%%%%%%%%%%%

\section{Introduction}

%%%%%%%%%%%%%%%%%%%%%%%%%%%%%%%%%%%%%%%%%%%%%%%%%%%%%%%%%%%%%%%%%%%%%%%%%%%%%%%

\subsection{EPR's elements of reality with projective measurements}

%%%%%%%%%%%%%%%%%%%%%%%%%%%%%%%%%%%%%%%%%%%%%%%%%%%%%%%%%%%%%%%%%%%%%%%%%%%%%%%

Bell~\cite{Bell64} discovered that some predictions of quantum
mechanics contradict Einstein, Podolsky, and Rosen's (EPR's)
``elements of reality''~\cite{EPR35}, defined as those satisfying
the following criterion: {\em ``If, without in any way disturbing
a system, we can predict with certainty (i.e., with probability
equal to unity) the value of a physical quantity, then there
exists an element of physical reality corresponding to this
physical quantity''}~\cite{comm}. In Bohm's version of the EPR's
experiment~\cite{Bohm51}, when two space-like separated projective
measurements~\cite{projective} of the spin along the same
direction $\vec n$ are performed on both spin-1/2 particles
prepared in the singlet state~\cite{notation}
\begin{equation}
|\psi\rangle={1 \over \sqrt{2}} (|n=+1, n=-1\rangle - |n=-1, n=+1\rangle),
\label{singlet}
\end{equation}
quantum theory predicts that the results ($+1$ or $-1$) will be
opposite for any $\vec n$. Therefore, there is an element of
reality corresponding to every spin component of either of the two
particles, since an observer performing a measurement of the spin
along $\vec n$ on particle 1 can predict with certainty the result
of a measurement of the spin along $\vec n$ on particle 2 without
disturbing it (since the particles are distant enough to exclude
communication at the speed of light or lower). In other words,
according to EPR, each particle must contain a set of instructions
\cite{Mermin81} which determines the result of any spin
measurement.

%%%%%%%%%%%%%%%%%%%%%%%%%%%%%%%%%%%%%%%%%%%%%%%%%%%%%%%%%%%%%%%%%%%%%%%%%%%%%%%

\subsection{Bell's inequality with projective measurements}

%%%%%%%%%%%%%%%%%%%%%%%%%%%%%%%%%%%%%%%%%%%%%%%%%%%%%%%%%%%%%%%%%%%%%%%%%%%%%%%

\begin{figure}
\centerline{\includegraphics[width=12cm]{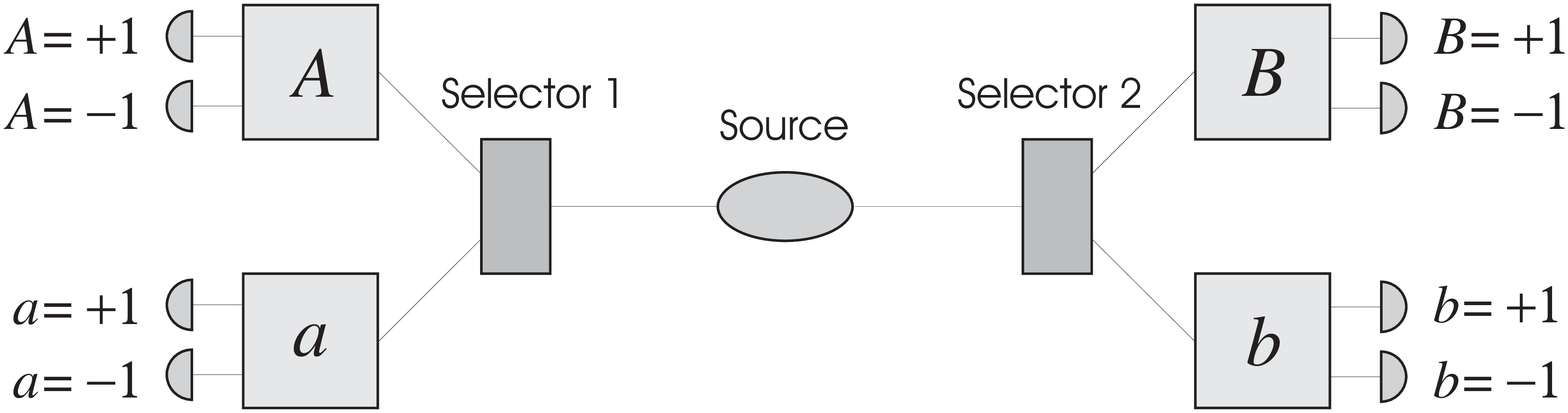}}
\caption{\label{stand} Standard configuration of an experiment to
test the CHSH inequality with two alternative projective
measurements on each particle, preceded by a device to randomly
select between them.}
\end{figure}

In the most common Bell's inequality, the
Clauser-Horne-Shimony-Holt (CHSH) inequality~\cite{CHSH69}, two
alternative dichotomic (i.e., with possible results $+1$ or $-1$)
observables $A$ or $a$ are measured on particle 1, and other two,
$B$ or $b$, on particle 2. For instance, these observables could
be the spin along $\vec A$ or $\vec a$ on particle 1 and along
$\vec B$ or $\vec b$ on particle 2. If the results of $A$, $a$,
$B$, $b$ are predefined for all pairs of particles, these results
must satisfy the CHSH inequality:
\begin{equation}
|\langle AB-Ab-aB-ab \rangle| \le 2,
\label{CHSH}
\end{equation}
where $\langle \, \rangle$ means average over all pairs. However,
in quantum mechanics two spin components of the same spin-1/2
particle, like $A$ and $a$ (or $B$ and $b$), cannot be measured in
the same experiment. Thus the quantum equivalent of the left-hand
side in inequality (\ref{CHSH}) is usually expressed as
\begin{equation}
\hat{B} = |\langle AB \rangle_\psi-\langle Ab \rangle_\psi-
\langle aB \rangle_\psi-\langle ab \rangle_\psi|,
\label{QCHSH}
\end{equation}
where $\langle AB \rangle_\psi$ is the mean value of the product
of the results of measuring~$A$ on particle~1 and $B$ on particle
2. It is a common assumption that any test of the CHSH inequality
requires two local observers who have free-will to choose between
two possible settings during the flight of the particles and thus
performing four different experiments ($AB$, $Ab$, $aB$, $ab$) on
four different subensembles of pairs (Fig.~\ref{stand}). Quantum
mechanics predict that, for pairs prepared in the singlet state
(\ref{singlet}), for certain choices of $\vec A$, $\vec a$, $\vec
B$, and $\vec b$, we will obtain $\hat{B} > 2$, which violates the
CHSH inequality (\ref{CHSH}). A widely extended belief is that a
setup with fixed local measurements cannot be used to test the
CHSH inequality.

In Sec.~\ref{Sec.I}, I will show that a suitable generalized
measurement is, from EPR's criterion of elements of reality,
equivalent to a random selection between two alternative
projective measurements. In Sec.~\ref{Sec.II}, I will show that a
violation of the CHSH inequality (\ref{CHSH}) can be obtained {\em
without} requiring local observers to choose between alternative
projective measurements.

%%%%%%%%%%%%%%%%%%%%%%%%%%%%%%%%%% Figure 1 %%%%%%%%%%%%%%%%%%%%%%%%%%%%%%%%%%%

%%%%%%%%%%%%%%%%%%%%%%%%%%%%%%%%%%%%%%%%%%%%%%%%%%%%%%%%%%%%%%%%%%%%%%%%%%%%%%%

\subsection{The locality and detection loopholes}

%%%%%%%%%%%%%%%%%%%%%%%%%%%%%%%%%%%%%%%%%%%%%%%%%%%%%%%%%%%%%%%%%%%%%%%%%%%%%%%

Experiments to test Bell's inequality
\cite{FC72,FT76,ADR82,SA88,OM88,OPKP92,TRO94,KMWZSS95,TBZG98,WJSHZ98,RKVSIMW01}
have agreed with quantum predictions and seem to exclude elements
of reality. However, up until now, all performed experiments are
subject to at least one of two loopholes.

The locality loophole~\cite{Bell81,Santos95} arises whenever
measurements performed on two spatially separated particles are
not space-like separated and thus the possibility of communication
at the speed of light between the two parts cannot be excluded.
The detection loophole~\cite{Pearle70,Santos92} arises from the
fact that in most experiments only a small subset of all the
created pairs are actually detected. It is therefore necessary to
assume that the registered pairs are a fair sample of all the
emitted pairs (fair sampling assumption). In practice, both
loopholes are not independent~\cite{GZ99}. These loopholes are
natural from the local hidden variables' point of view, in which
particles have additional hidden variables that enable them to
give results for certain experiments and not for others (for
instance, to pass an analyzer for certain settings and not for
others). If the actual setting does not correspond to the hidden
variables of the particle, then, according to the detection
loophole, the particle is not detected. Whereas, according to the
locality loophole, this situation never happens because the
particle knows the setting in advance.

The experiment in~\cite{WJSHZ98} with polarization-entangled
photons and a $400$ m separation between the particles (which
gives the observer $1.3$ $\mu$s to make the selection-measurement
process, defined in~\cite{WJSHZ98} ``to last from the first point
in time which can influence the choice of the analyzer setting
until the final registration of the photon'') is not subject to
the locality loophole, but the detection efficiency ($5\%$) is not
high enough to close the detection loophole ($82.8\%$ would be
required~\cite{CH74}).

On the other hand, the experiment in~\cite{RKVSIMW01} with trapped
beryllium ions has nearly perfect detection efficiency and thereby
is not subject to the detection loophole, but the distance between
the ions ($3\,\mu$m), although large enough that no known
interaction could affect the results, is not large enough to close
the locality loophole, because the selection-measurement requires
two steps: a selection (equivalent to rotating a wave-plate in the
case of experiments with polarized photons) applying Raman beams
for a pulse of a duration of~$400$~ns, and a measurement probing
the ion with circularly polarized light from a ``detector'' laser
beam during this detection pulse; if the ion is in one state, it
scatters many photons; if it is in the orthogonal state, it
scatters very few photons.

It was first thought that improving the detection efficiency in
experiments with pairs of entangled photons would avoid both
loopholes, but this proved more difficult than expected and,
despite several proposals~\cite{KESC94,FFS95}, no conclusive
experiment has been achieved. Another approach based on pairs of
atoms produced through a photodissociation process
\cite{LS81,FWL95}, or pairs of Rydberg atoms~\cite{FAHS96}, is not
easy to implement and no conclusive test of Bell's inequality has
been carried out in these systems~\cite{VXK02}.

Summing up, although the recent experiments
\cite{WJSHZ98,RKVSIMW01} have meant a significant advance, they
still have not settled the debate~\cite{Vaidman01}. A
loophole-free experiment is still demanded
\cite{Vaidman01,Santos01}.

In Sec.~\ref{Sec.III} we will show that the approach to test
Bell's inequality introduced in Secs.~\ref{Sec.I} and \ref{Sec.II}
could be useful to reduce the distance requirements to close the
locality loophole in experiments with a high enough detection
efficiency.

%%%%%%%%%%%%%%%%%%%%%%%%%%%%%%%%%% Figure 2 %%%%%%%%%%%%%%%%%%%%%%%%%%%%%%%%%%%

\begin{figure}
\centerline{\includegraphics[width=12cm]{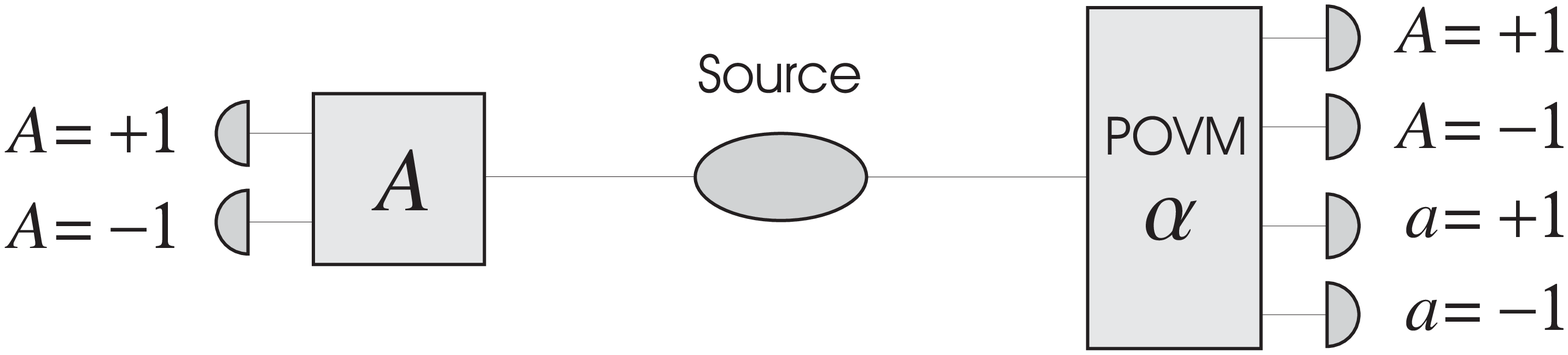}}
\caption{\label{modf} Modified EPR-Bohm experiment: the
measurement on the left particle is a projective measurement,
while the measurement on the right particle is a generalized
measurement described by the POVM (\ref{POVM1})-(\ref{POVM4}).}
\end{figure}

%%%%%%%%%%%%%%%%%%%%%%%%%%%%%%%%%%%%%%%%%%%%%%%%%%%%%%%%%%%%%%%%%%%%%%%%%%%%%%%

\section{EPR's elements of reality with POVMs}
\label{Sec.I}

%%%%%%%%%%%%%%%%%%%%%%%%%%%%%%%%%%%%%%%%%%%%%%%%%%%%%%%%%%%%%%%%%%%%%%%%%%%%%%%

Motivated by the quantum information approach to quantum mechanics
and by the fact that current technology allows an exquisite level
of control over the measurements that can be performed, recent
formulations of the principles of quantum mechanics
\cite{Peres93,NC00} stress that the measurements correspond to
positive operator-valued measures (POVMs)
\cite{Ludwig76,Davies76,Peres93,NC00}, extending the notion of von
Neumann's projection-valued measures. The main difference between
POVMs and von Neumann's projection-valued measures is that for
POVMs the number of available outcomes of a measurement may be
higher than the dimensionality of the Hilbert space. An
$N$-outcome generalized measurement is represented by an
$N$-element POVM which consists of $N$ positive-semidefinite
operators $\{E_d\}$ that sum the identity (i.e., $\sum_d E_d=
1\!\!\:\!{\rm{I}}$). Neumark's theorem~\cite{Neumark43} guarantees
that there always exists a realizable experimental procedure to
generate any desired POVM. Any generalized measurement represented
by a POVM can be seen as a von Neumann's measurement on a larger
Hilbert space. Therefore, any generalized measurement on a single
qubit can be seen as a von Neumann's joint measurement on a system
composed by the qubit plus an auxiliary quantum system (ancilla).

A natural question is what POVMs means from the point of view of
EPR's elements of reality. To answer that, let us go back to the
EPR-Bohm experiment. Let us suppose that, instead of the same
projective measurement on both particles, a projective measurement
$A$ is performed on particle 1 and a 4-outcome generalized
measurement is performed on particle 2 (Fig.~\ref{modf}).
Specifically, let us consider the generalized measurement
represented by the following 4-element POVM:
\begin{eqnarray}
E_{A+} & = & {1 \over 2} P_{\neg |A=-1\rangle} \nonumber \\ & = &
{1 \over 2} \left(1\!\!\:\!{\rm{I}}-P_{|A=-1\rangle}\right),
\nonumber \\
& = & {1 \over 2} |A=+1\rangle\langle A=+1|,
\label{POVM1} \\
E_{A-} & = & {1 \over 2} P_{\neg |A=+1\rangle} \nonumber \\ & = &
{1 \over 2} \left(1\!\!\:\!{\rm{I}}-P_{|A=+1\rangle}\right),
\nonumber \\
& = & {1 \over 2} |A=-1\rangle\langle A=-1|,
\label{POVM2} \\
E_{a+} & = & {1 \over 2} P_{\neg |a=-1\rangle} \nonumber \\ & = &
{1 \over 2} \left(1\!\!\:\!{\rm{I}}-P_{|a=-1\rangle}\right),
\nonumber \\
& = & {1 \over 2} |a=+1\rangle\langle a=+1|,
\label{POVM3} \\
E_{a-} & = & {1 \over 2} P_{\neg |a=+1\rangle} \nonumber \\ & = &
{1 \over 2} \left(1\!\!\:\!{\rm{I}}-P_{|a=+1\rangle}\right),
\nonumber \\
& = & {1 \over 2} |a=-1\rangle\langle a=-1|,
\label{POVM4}
\end{eqnarray}
where $P_{\neg |A=-1\rangle}$ is the projection on states
orthogonal to $|A=-1\rangle$. If the result of the POVM is that
corresponding to $E_{A+}$, this means that the initial state was
{\em not} $|A=-1\rangle$~\cite{Ivanovic87,Peres88}. From the point
of view of EPR's criterion of elements of reality, both $A$ and
$a$ must have predefined values $+1$ or $-1$; thus any measurement
which reveals that $A$ was not $-1$ also reveals that $A$ was
$+1$. Therefore, following EPR's point of view, we can label the
corresponding output of the POVM as $A=+1$, and likewise for the
other three possible outcomes, as in Fig.~\ref{modf}.

However, not only $A$ but also $a$ must have elements of reality.
Why then do we obtain only one of them after a POVM? To answer
this question, it is useful to keep in mind the fact that any
desired POVM with a finite number of elements can be converted
into a projective measurement by introducing an auxiliary,
independently prepared, quantum system (ancilla)
\cite{Peres93,NC00,Neumark43}. The POVM can then be seen as a
projective measurement on the system composed by the original
particle and the ancilla. One way of implementing the POVM given
by (\ref{POVM1})-(\ref{POVM4}) is by measuring the observable
\begin{eqnarray}
\hat{O} & = & r_{A+} P_{|A=+1, z=+1\rangle}+r_{A-} P_{|A=-1, z=+1\rangle} \nonumber \\
& & + r_{a+} P_{|a=+1, z=-1\rangle}+r_{a-} P_{|a=-1, z=-1\rangle},
\end{eqnarray}
where $P_{|A=+1, z=+1\rangle}$ is the projector onto state $| A=+1
\rangle$ of the particle and state $| z=+1 \rangle$ of the
ancilla, and $r_{A+}$ is the corresponding result. One of the
possible ways to measure $\hat{O}$ is by preparing the ancilla in
the maximally mixed state $\rho = {1 \over 2} 1\!\!\:\!{\rm{I}}$,
then measuring $z$ on the ancilla and then measuring $A$ (if the
result of the previous measurement is $z=+1$) or $a$ (if the
result is $z=-1$) on the particle. Such a procedure is analogous
to the one followed in a standard test of Bell's inequality with
alternative projective measurements. The result of the first
measurement acts as a random generator (the two possible outcomes
are unpredictable and have the same probability of occurring)
which determines the projective measurement that is finally
chosen. Therefore, a measurement of $\hat{O}$
%and thus the POVM (\ref{POVM1})-(\ref{POVM4}),
is equivalent to a selection between $A$ and $a$ using the result
of a projective measurement on the ancilla, followed by a
projective measurement of either $A$ or $a$ on the particle. The
randomness is provided by a quantum measurement on the ancilla.
The result of this measurement selects one experiment or other on
the particle. Therefore, we conclude that the described
implementation of the POVM (\ref{POVM1})-(\ref{POVM4}) on particle
1 of a two-qubit system in the singlet state is, from the point of
view of EPR's criterion of elements of reality, equivalent to two
alternative dichotomic projective measurements $A$ or $a$ preceded
by a device to randomly choose between them.

Let us assume that every implementation of a projective
measurement is equivalent. Then, the POVM
(\ref{POVM1})-(\ref{POVM4}) can be equivalently measured by a {\em
single} projective measurement of the observable $\hat{O}$ on the
particle-ancilla system. In this case, the usual
selection-measurement process in each of the wings of a test of
Bell's inequality is replaced with a single measurement on a
particle-ancilla system.

%%%%%%%%%%%%%%%%%%%%%%%%%%%%%%%%%% Figure 3 %%%%%%%%%%%%%%%%%%%%%%%%%%%%%%%%%%%

\begin{figure}
\centerline{\includegraphics[width=12cm]{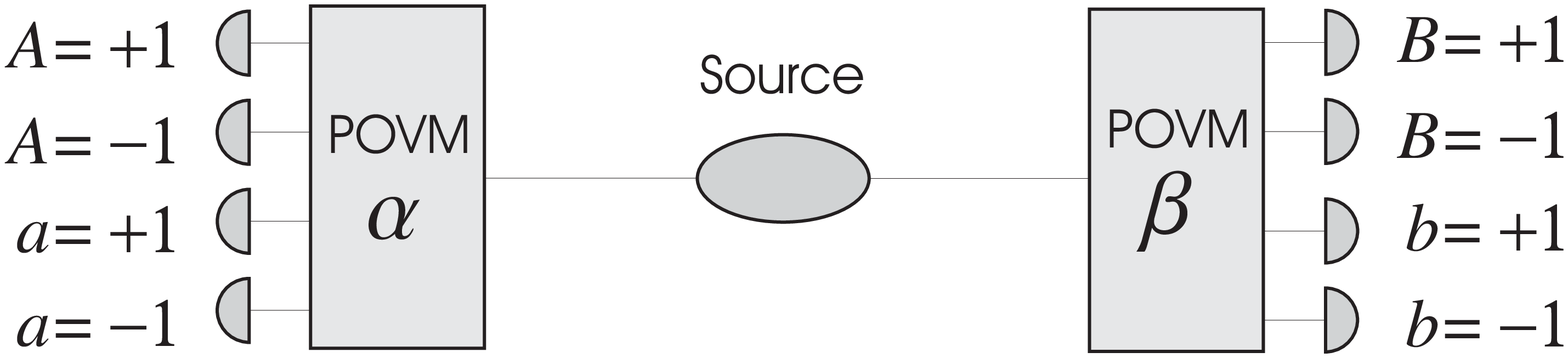}}
\caption{\label{prop} Proposed configuration of an experiment to
test the CHSH inequality without alternative settings.}
\end{figure}

%%%%%%%%%%%%%%%%%%%%%%%%%%%%%%%%%%%%%%%%%%%%%%%%%%%%%%%%%%%%%%%%%%%%%%%%%%%%%%%

\section{Violating Bell's inequality with POVMs}
\label{Sec.II}

%%%%%%%%%%%%%%%%%%%%%%%%%%%%%%%%%%%%%%%%%%%%%%%%%%%%%%%%%%%%%%%%%%%%%%%%%%%%%%%

The next step is to show that the predictions of quantum mechanics
for a singlet state violate the CHSH inequality (\ref{CHSH}) when
each local observer measures a POVM of the type
(\ref{POVM1})-(\ref{POVM4}). Let us suppose that the POVM $\alpha$
defined in~(\ref{POVM1})-(\ref{POVM4}) is measured on particle~1,
and a similar POVM $\beta$, defined as in
(\ref{POVM1})-(\ref{POVM4}) just by changing $A$ by $B$ and $a$ by
$b$, is measured on particle 2 (Fig.~\ref{prop}). The average~$AB$
(and similarly the other three) appearing in the CHSH
inequality~(\ref{CHSH}) can be calculated as follows:
\begin{eqnarray}
\langle AB \rangle_\psi & = & [P_\psi (E_{A+},E_{B+}) -
P_\psi (E_{A+},E_{B-}) \nonumber \\ & &
- P_\psi (E_{A-},E_{B+})
+ P_\psi (E_{A-},E_{B-})] / \nonumber \\ & &
[P_\psi (E_{A+},E_{B+}) + P_\psi (E_{A+},E_{B-}) \nonumber \\ & &
+ P_\psi (E_{A-},E_{B+}) + P_\psi (E_{A-},E_{B-})],
\end{eqnarray}
where $P_\psi (E_{A+},E_{B+})$ is the probability of the observer
of particle 1 obtaining $A=+1$, and the observer of particle 2
obtaining $B=+1$. The denominator is the probability of the result
of the POVM on particle 1 giving a result belonging to $\{
E_{A+},E_{A-}\}$, and the result of the POVM on particle 2 giving
a result belonging to $\{E_{B+},E_{B-}\}$. Probabilities for the
outcomes obey the Born rule for POVMs:
\begin{equation}
P_\psi (E_{A+},E_{B+}) = \langle\psi| E_{A+} \otimes E_{B+} |\psi\rangle.
\end{equation}
Therefore, it is easy to see that, for the singlet state (\ref{singlet}),
\begin{equation}
\langle AB \rangle_\psi = - \cos \theta_{AB},
\end{equation}
where $\theta_{AB}$ is the angle between $\vec A$ and $\vec B$.
Choosing $\vec A = (1,0,0)$, $\vec a = (0,0,1)$, $\vec
B=(-1,0,1)/\sqrt{2}$, and $\vec b =(1,0,1)/\sqrt{2}$, we obtain $\hat{
B} = 2\sqrt{2}$, which violates the CHSH inequality (\ref{CHSH}).
Thus, we conclude that the predictions of quantum mechanics for
the singlet state violate the CHSH inequality, even if local
observers do not have to choose or switch between alternative
projective measurements; a suitable POVM implements such a
selection-measurement process in a single step.

%%%%%%%%%%%%%%%%%%%%%%%%%%%%%%%%%%%%%%%%%%%%%%%%%%%%%%%%%%%%%%%%%%%%%%%%%%%%%%%

\section{POVMs and loophole-free tests}
\label{Sec.III}

%%%%%%%%%%%%%%%%%%%%%%%%%%%%%%%%%%%%%%%%%%%%%%%%%%%%%%%%%%%%%%%%%%%%%%%%%%%%%%%

\subsection{Assumption on the ability of local ``hidden'' variables}

%%%%%%%%%%%%%%%%%%%%%%%%%%%%%%%%%%%%%%%%%%%%%%%%%%%%%%%%%%%%%%%%%%%%%%%%%%%%%%%

To avoid the locality loophole, Bell stressed the importance of
experiments ``in which the settings are changed during the flight
of the particles''~\cite{Bell81}. Contrary to what happens in
experiments with passive switches~\cite{TBZG98,GZ99}, in the POVM
approach one could not reasonably argue that any particle of the
pair could guess the ``settings''. One might think that particle 1
could know the hidden-variable state of the ancilla associated to
particle 2 and that this state could determine the ``setting'' on
particle 2. However, a similar complaint might be applied to any
standard test of the CHSH inequality, only changing
``hidden-variable state of the ancilla'' to ``hidden-variable
state of the acousto-optical switch'' (or any other supposedly
``random'' method for switching). In any test of Bell's inequality
we have to assume that there are some limits to the ability of the
hidden variables of particle 2 to ascertain what is going to be
done to particle 1; otherwise the locality loophole could not be
avoided by any conceivable setup.

Explicitly, our approach hinges on an additional assumption: $(h)$
The particle on which POVM $\alpha$ is going to be measured has no
ability to ascertain or dictate whether the element of reality,
$A$ or $a$, is going to be revealed. Such assumption is equivalent
to the following usual assumption in the standard approach: $(h')$
The local hidden variables have no capability to ascertain or
dictate in advance the final setting of the random mechanism.

Supporting the equivalence of assumptions $(h)$ and $(h')$ is the
following argument: The POVM can be implemented as a single and
projective measurement on the particle-ancilla system. This
implies an interaction between a measurement apparatus (of which
the ancilla is a part) and the particle. From the perspective of
the elements of reality, the result of this interaction has two
meanings: on one hand, it reveals a pre-existent value; on the
other, it entails a choice. Following EPR, the former can be
regarded as an element of reality. However, nothing in the EPR
criterion authorizes us to suppose that this ``choice'' is somehow
pre-existent neither in the particle nor in the apparatus-ancilla
system. The ``choice'' is not an EPR element of reality. It is the
result of an essentially uncontrollable, intrinsically
unrepeatable, and basically unpredictable phenomenon which
involves interactions between the particle, the apparatus, the
ancilla, and the environment in a particular region of space-time.
%The ``choice'' is ideally simultaneous to the ``revelation'' of an
%element of reality.
A phenomenon which, to our present knowledge, can be used as a
source of randomness~\cite{SGGGZ00,JAWWZ00} which offers the
advantage over conventional randomness sources (like electronic
noise) of being robust and invulnerable to environmental
perturbations. Why then not judge assumption $(h)$ to be at least
as limiting, if not less, than the usual assumption $(h')$?

%%%%%%%%%%%%%%%%%%%%%%%%%%%%%%%%%%%%%%%%%%%%%%%%%%%%%%%%%%%%%%%%%%%%%%%%%%%%%%%

\subsection{Practical advantages}

%%%%%%%%%%%%%%%%%%%%%%%%%%%%%%%%%%%%%%%%%%%%%%%%%%%%%%%%%%%%%%%%%%%%%%%%%%%%%%%

At first sight, the standard approach, based on a random selection
between two alternative projective measurements, and the approach
introduced here, based on single POVM implemented by a single
projective measurement on the particle-ancilla system, might be
considered to be physically equivalent. However, the equivalence
is not complete. The difference could be useful (or at least must
be taken into account) in designing loophole-free tests of Bell's
inequality.

Detection efficiencies that are high enough to avoid the detection
loophole are usually associated to entangled pairs of massive
particles. However, preparing and preserving entanglement between
two distant massive particles is a difficult task. Therefore, an
interesting question is: which is the {\em minimum} distance
between the local measurements on the particles necessary to avoid
the locality loophole? This distance is $ct$, where $c$ is the
speed of light in a vacuum. In the standard approach, $t=t_S+t_M$,
where $t_S$ is the time which elapses since the moment in which
the final setting of the selector {\em can be ascertained} until
the particle enters the measurement device, and $t_M$ is the time
that the projective measurement lasts. In the POVM approach, it is
just $t=t_M$.

Moreover, while $t_S$ cannot be neglected because it involves a
sequence of physical processes (it requires, among other things,
connecting a mechanism to generate random results with another
which fixes the selector), $t_M$ is, in principle, comparatively
smaller. The behavior of a quantum system subject to a projective
measurement is described by von Neumann's reduction
postulate~\cite{vonNeumann52}. In this description, measurement is
{\em instantaneous}. However, real measurements require
interactions between the system being measured and the
environment, interactions which are not actually instantaneous.
However, for our discussion, the important point is that there is
no fundamental limit to the minimum time required for a
measurement. Therefore, in the POVM approach to test Bell's
inequality, {\em in principle}, there would be no limit to the
minimum spatial distance between local measurements. In practice,
this means that, in the POVM approach, this distance is only
restricted by the duration of a single projective
measurement~\cite{comm2}.

%%%%%%%%%%%%%%%%%%%%%%%%%%%%%%%%%%%%%%%%%%%%%%%%%%%%%%%%%%%%%%%%%%%%%%%%%%%%%%%

\section{Conclusions}

%%%%%%%%%%%%%%%%%%%%%%%%%%%%%%%%%%%%%%%%%%%%%%%%%%%%%%%%%%%%%%%%%%%%%%%%%%%%%%%

I have proposed a different approach to test the Bell-CHSH
inequality. It basically consists in replacing each of the usual
pairs of alternative projective measurements preceded by a random
mechanism to select between them, by a single fixed POVM. The
basic assumption tested by the proposed experiment is the
existence of elements of reality as defined by EPR. In
Sec.~\ref{Sec.I}, it has been shown that the POVM $\alpha$ reveals
the value of an EPR element of reality, either $A$ or $a$. Since
the CHSH inequality is derived from the assumption of the
existence of EPR elements of reality, the violation of the CHSH
inequality showed in Sec.~\ref{Sec.II} can be interpreted as a
test of the nonexistence of EPR elements of reality.

However, the argument hinges on an additional assumption: $(h)$
the particle on which the POVM $\alpha$ is going to be measured
has no capability to ascertain or dictate which of the element of
reality, $A$ or $a$, is going to be revealed. In
Sec.~\ref{Sec.III}, it is argued that such an assumption is
equivalent to the usual assumption in the standard approach:
$(h')$ the local hidden variables have no capability to ascertain
or dictate in advance the final setting of the random mechanism.
Any test of Bell's inequality requires an assumption of this kind.
It is stressed that $(h)$ imposes the same or fewer restrictions
than $(h')$.

Finally, the advantages of this approach are twofold: (a) it does
not require us to assume observers having ``free-will'', and (b)
it can potentially be used to improve the perspectives of success
in the race for a loophole-free test of Bell's inequality. This is
because the new approach eliminates the requirement of a
``random'' external mechanism before a projective measurement and
replaces both the random mechanism and the projective measurement
itself with a single projective measurement. For some physical
systems, at least, this implies a reduction of the spatial length
between the local measurements needed to avoid the locality
loophole. Such a reduction is of practical interest, since the
typical physical systems which allow us to avoid the detection
loophole (namely, entangled massive particles) do not allow a
significant separation among parts.

%%%%%%%%%%%%%%%%%%%%%%%%%%%%%% Acknowledgements %%%%%%%%%%%%%%%%%%%%%%%%%%%%%%%

\section*{Acknowledgments}
I thank J. Calsamiglia, J. L. Cereceda, J. I. Cirac, J.
Finkelstein, L. Masanes, E. Santos, J. A. Smolin, and H.
Weinfurter for illuminating discussions, J. Calsamiglia for his
hospitality at Harvard University, H. Weinfurter for his
hospitality at Ludwig-Maximilians-Universit\"{a}t, M\"{u}nchen,
and Max-Planck-Institut f\"{u}r Quantenoptik, Garching, and the
Spanish Ministerio de Ciencia y Tecnolog\'{\i}a grants
BFM2001-3943 and BFM2002-02815, the Junta de Andaluc\'{\i}a grant
FQM-239, and the Max-Planck-Institut f\"{u}r Quantenoptik for
support.

%%%%%%%%%%%%%%%%%%%%%%%%%%%%%%%%%% References %%%%%%%%%%%%%%%%%%%%%%%%%%%%%%%%%%


\begin{thebibliography}{00}

\bibitem{Bell64}
J.S. Bell,
%``On the Einstein-Podolsky-Rosen paradox'',
Physics (Long Island City, NY) {\bf 1}, 195 (1964).

\bibitem{EPR35}
A. Einstein, B. Podolsky, and N. Rosen,
%``Can quantum-mechanical description of physical reality
%be considered complete?'',
Phys. Rev. {\bf 47}, 777 (1935).

\bibitem{comm}
Locality assumes that every local measurement is determined by
local hidden variables. EPR do not assume that all local
measurements have pre-determined results, except for those which
satisfy the criterion of elements of reality. From this point of
view, EPR theories are just a subset of local hidden-variable
theories.

\bibitem{Bohm51}
D. Bohm,
{\em Quantum Theory}
(Prentice-Hall, Englewood Cliffs, NJ, 1951), p. 611.

\bibitem{projective}
Projective measurements are those represented in quantum mechanics
by self-adjoint operators $O=\sum_{r} r P_r$, where $P_r$ is the
projector onto the eigenspace of~$O$ with eigenvalue $r$. The
possible outcomes of the measurement correspond to the eigenvalues
$r$. $\{P_r\}$ are mutually orthogonal projectors.

\bibitem{notation}
$|n=+1, n=-1\rangle$ represents the quantum state in which, when
the spin along~$\vec n$ is measured on both particles, we obtain
the result $+1$ for particle 1, and $-1$ for particle 2.

\bibitem{Mermin81}
N.D. Mermin,
%``Bringing home the atomic world: Quantum mysteries for anyone'',
%{\em Am. J. Phys.} {\bf 49}, 940-943 (1981).
Am. J. Phys. {\bf 49}, 940 (1981).

\bibitem{CHSH69}
J.F. Clauser, M.A. Horne, A. Shimony, and R.A. Holt,
%``Proposed experiment to test local hidden-variable theories'',
Phys. Rev. Lett. {\bf 23}, 880 (1969).

%%%%%%%%%%%%%%%%%%%%%%%%%%%%%%%%% Experiments %%%%%%%%%%%%%%%%%%%%%%%%%%%%%%%%%

\bibitem{FC72}
S.J. Freedman and J.F. Clauser,
%``Experimental test of local hidden-variable theories'',
%{\em Phys. Rev. Lett.} {\bf 28}, 14, 938-941 (1972).
Phys. Rev. Lett. {\bf 28}, 938 (1972).

\bibitem{FT76}
E.S. Fry and R.C. Thompson,
%``Experimental test of local hidden-variable theories'',
%{\em Phys. Rev. Lett.} {\bf 37}, 8, 465-468 (1976).
Phys. Rev. Lett. {\bf 37}, 465 (1976).

\bibitem{ADR82}
A. Aspect, J. Dalibard, and G. Roger,
%``Experimental test of Bell's inequalities using time-varying
%analyzers'',
Phys. Rev. Lett. {\bf 49}, 1804 (1982).

\bibitem{SA88}
Y.H. Shih and C.O. Alley,
%``New type of Einstein-Podolsky-Rosen-Bohm experiment using pairs
%of light quanta produced by optical parametric down conversion'',
%{\em Phys. Rev. Lett.} {\bf 61}, 26, 2921-2924 (1988).
Phys. Rev. Lett. {\bf 61}, 2921 (1988).

\bibitem{OM88}
Z.Y. Ou and L. Mandel,
%``Violation of Bell's inequality
%and classical probability in a two-photon correlation experiment'',
%{\em Phys. Rev. Lett.} {\bf 61}, 1, 50-53 (1988).
Phys. Rev. Lett. {\bf 61}, 50 (1988).

\bibitem{OPKP92}
Z.Y. Ou, S.F. Pereira, H.J. Kimble, and K.C. Peng,
%``Realization of the Einstein-Podolsky-Rosen paradox for continuous
%variables'',
%{\em Phys. Rev. Lett.} {\bf 68}, 25, 3663-3666 (1992).
Phys. Rev. Lett. {\bf 68}, 3663 (1992).

\bibitem{TRO94}
P.R. Tapster, J.G. Rarity, and P.C.M. Owens,
%``Violation of Bell's inequality over 4 km of optical fiber'',
%{\em Phys. Rev. Lett.} {\bf 73}, 14, 1923-1926 (1994).
Phys. Rev. Lett. {\bf 73}, 1923 (1994).

\bibitem{KMWZSS95}
P.G. Kwiat, K. Mattle, H. Weinfurter, A. Zeilinger, A.V.
Sergienko, and Y.H. Shih,
%``New high-intensity source of polarization-entangled photon pairs'',
%{\em Phys. Rev. Lett.} {\bf 75}, 24, 4337-4341 (1995).
Phys. Rev. Lett. {\bf 75}, 4337 (1995).

\bibitem{TBZG98}
W. Tittel, J. Brendel, H. Zbinden, and N. Gisin,
%``Violation of Bell inequalities by photons more than 10 km apart'',
%{\em Phys. Rev. Lett.} {\bf 81}, 17, 3563-3566 (1998);
Phys. Rev. Lett. {\bf 81}, 3563 (1998).

\bibitem{WJSHZ98}
G. Weihs, T. Jennewein, C. Simon, H. Weinfurter,
and A. Zeilinger,
%``Violation of Bell's inequality under strict
%Einstein locality conditions'',
Phys. Rev. Lett. {\bf 81}, 5039 (1998).

\bibitem{RKVSIMW01}
M.A. Rowe, D. Kielpinski, V. Meyer,
C.A. Sackett, W.M. Itano, C. Monroe, and D.J. Wineland,
%``Experimental violation of a Bell's inequality
%with efficient detection'',
Nature (London) {\bf 409}, 791 (2001).

%%%%%%%%%%%%%%%%%%%%%%%%%%%%%%%%%% Loopholes %%%%%%%%%%%%%%%%%%%%%%%%%%%%%%%%%%

\bibitem{Bell81}
J.S. Bell,
%``Bertlmann's socks and the nature of reality'',
%{\em Journal de Physique}, Colloque {\em C2},
%suppl. au numero 3, Tome 42, 41-61 (1981).
J. Phys. C (Paris) {\bf 2}, 41 (1981).

\bibitem{Santos95}
E. Santos,
%``Constraints for the violation of the Bell
%inequality in Einstein-Podolsky-Rosen-Bohm experiments'',
%{\em Phys. Lett. A} {\bf 200}, 1, 1-6 (1995).
Phys. Lett. A {\bf 200}, 1 (1995).

\bibitem{Pearle70}
P.M. Pearle,
%``Hidden-variable example based upon data rejection'',
%{\em Phys. Rev. D} {\bf 2}, 8, 1418-1425 (1970).
Phys. Rev. D {\bf 2}, 1418 (1970).

\bibitem{Santos92}
E. Santos,
%``Critical analysis of the empirical tests of local
%hidden-variable theories'',
%{\em Phys. Rev. A} {\bf 46}, 7, 3646-3656 (1992).
Phys. Rev. A {\bf 46}, 3646 (1992).

\bibitem{GZ99}
N. Gisin and H. Zbinden,
%``Bell inequality and the locality loophole: Active versus passive switches'',
%{\em Phys. Lett. A} {\bf 264}, 2-3, 103-107 (1999).
Phys. Lett. A {\bf 264}, 103 (1999).

\bibitem{CH74}
J.F. Clauser and M.A. Horne,
%``Experimental consecuences of objective local theories'',
Phys. Rev. D {\bf 10}, 526 (1974).

%%%%%%%%%%%%%%%%%%%%%%%%%%% Loophole-free proposals %%%%%%%%%%%%%%%%%%%%%%%%%%%

\bibitem{KESC94}
P.G. Kwiat, P.H. Eberhard, A.M. Steinberg, and R.Y. Chiao,
%``Proporsal for a loophole-free Bell inequality experiment'',
%{\em Phys. Rev. A} {\bf 49}, 5, Part A, 3209-3220 (1994).
Phys. Rev. A {\bf 49}, 3209 (1994).

\bibitem{FFS95}
S.F. Huelga, M. Ferrero, and E. Santos,
%S.G. Fern\'{a}ndez Huelga, M. Ferrero, and E. Santos,
%``Loophole-free test of the Bell inequality'',
%{\em Phys. Rev. A} {\bf 51}, 6, 5008-5011 (1995).
Phys. Rev. A {\bf 51}, 5008 (1995).

\bibitem{LS81}
T.K. Lo and A. Shimony,
%``Proposed molecular test of local hidden-variables theories'',
%{\em Phys. Rev. A} {\bf 23}, 6, 3003-3012 (1981).
Phys. Rev. A {\bf 23}, 3003 (1981).

\bibitem{FWL95}
E.S. Fry, T. Walther, and S. Li,
%``Proposal for a loophole-free test of the Bell inequalities'',
%{\em Phys. Rev. A} {\bf 52}, 6, 4381-4395 (1995).
Phys. Rev. A {\bf 52}, 4381 (1995).

\bibitem{FAHS96}
M. Freyberger, P.K. Aravind, M.A. Horne, and A. Shimony,
%``Proposed test of Bell's inequality without
%detection loophole by using entangled Rydberg atoms'',
%{\em Phys. Rev. A} {\bf 53}, 3, 1232-1244 (1996).
Phys. Rev. A {\bf 53}, 1232 (1996).

\bibitem{VXK02}
Another method has recently become available for producing
correlated atoms, namely non-linear mixing in Bose-Einstein
condensates: J.M. Vogels, K. Xu, and W. Ketterle,
%``Generation of macroscopic pair-correlated atomic
%beams by four wave mixing in Bose-Einstein condensates'',
%{\em Phys. Rev. Lett.} {\bf 89}, 2, 020401 (2002).
Phys. Rev. Lett. {\bf 89}, 020401 (2002).

%%%%%%%%%%%%%%%%%%%%%%%%%%%%%%%%% Criticisms %%%%%%%%%%%%%%%%%%%%%%%%%%%%%%%%%%

\bibitem{Vaidman01}
L. Vaidman,
%``Tests of Bell inequalities'',
Phys. Lett. A {\bf 286}, 241 (2001).

\bibitem{Santos01}
E. Santos,
%``Quantum mechanics vs. local realism, is that the question?'',
quant-ph/0103062.

%%%%%%%%%%%%%%%%%%%%%%%%%%%%%%%%%%%% POVMs %%%%%%%%%%%%%%%%%%%%%%%%%%%%%%%%%%%%

\bibitem{Peres93}
A. Peres,
{\em Quantum Theory: Concepts and Methods} (Kluwer, Dordrecht,
1993).

\bibitem{NC00}
M.A. Nielsen and I.L. Chuang,
{\em Quantum Computation and Quantum Information}
(Cambridge University Press, Cambridge, 2000).

\bibitem{Ludwig76}
G. Ludwig,
{\em Einf\"{u}hrung in die Grundlagen der Theoretischen Physik}
(Vieweg, Braunschweig, 1976).

\bibitem{Davies76}
E.B. Davies,
{\em Quantum Theory of Open Systems}
(Academic Press, New York, 1976).

\bibitem{Ivanovic87}
I.D. Ivanovic,
%``How to differenciate between non-orthogonal states'',
%{\em Phys. Lett. A} {\bf 123}, 6, 257-259 (1987).
Phys. Lett. A {\bf 123}, 257 (1987).

\bibitem{Peres88}
A. Peres,
%``How to differentiate between non-orthogonal states'',
%{\em Phys. Lett. A} {\bf 128}, 19-24 (1988).
Phys. Lett. A {\bf 128}, 19 (1988).

\bibitem{Neumark43}
M.A. Neumark,
%``?'',
%C. R. Acad. Sci. URSS {\bf 41}, 359-? (1943);
C. R. Acad. Sci. URSS {\bf 41}, 359 (1943).
%``Operatorenalgebren im Hilbertschen Raum'',
%in {\em Sowjetische Arbeiten zur Funktionalanalysis}, Verlag Kultur und
%Fortschritt, Berlin, 1954.

%%%%%%%%%%%%%%%%%%%%%%%%%% Random number generators %%%%%%%%%%%%%%%%%%%%%%%%%%%

\bibitem{SGGGZ00}
A. Stefanov, N. Gisin, O. Guinnard, L. Guinnard, and H. Zbinden,
%``Optical quantum random number generator'',
%{\em J. Mod. Opt.} {\bf 47}, 4, 595-598 (2000).
J. Mod. Opt. {\bf 47}, 595 (2000).

\bibitem{JAWWZ00}
T. Jennewein, U. Achleitner, G. Weihs, H. Weinfurter, and A. Zeilinger,
%``A fast and compact quantum random number generator'',
%{\em Rev. Sci. Instrum.} {\bf 71}, 1675–1680 (2000).
Rev. Sci. Instrum. {\bf 71}, 1675 (2000).

%%%%%%%%%%%%%%%%%%%%%%%%%%%%%%%%%%%%%%%%%%%%%%%%%%%%%%%%%%%%%%%%%%%%%%%%%%%%%%%

\bibitem{vonNeumann52}
J. von Neumann,
{\em Mathematische Grundlagen der Quantenmechanik}
(Springer-Verlag, Berlin, 1932).
%English version: {\em Mathematical Foundations of Quantum Mechanics},
%(Princeton University Press, Princeton, NJ, 1955).

\bibitem{comm2}
We have assumed that every projective measurement on a composite
system can be implemented by a single step process. However, POVMs
are sometimes implemented as a two-step process (like, for
instance, in the case of ions) and thus no time would be gained.

%%%%%%%%%%%%%%%%%%%%%%%%%%%%%%%%%%%%%%%%%%%%%%%%%%%%%%%%%%%%%%%%%%%%%%%%%%%%%%%

\end{thebibliography}
\end{document}